\documentclass[prb,aps,twocolumn,showpacs,amsmath,amssymb]{revtex4}
\usepackage{amsmath,graphicx,dcolumn}

\newcommand{\CaFeAs}{CaFe$_2$As$_2$}

\newcommand{\BaNiAs}{BaNi$_2$As$_2$}

\newcommand{\BaNiP}{BaNi$_2$P$_2$}
\newcommand{\ThCrSi}{ThCr$_2$Si$_2$}
\newcommand{\Sr}{SrNi$_2$P$_2$}

\begin{document}

\title{Superconductivity and the Effects of Pressure and Structure in Single Crystaline SrNi$_2$P$_2$}

\author{F. Ronning$^1$, E.D. Bauer$^1$, T. Park$^{1,2}$, S.-H. Baek$^1$, H. Sakai$^{1,3}$, J.D. Thompson$^1$}
\affiliation{$^1$Los Alamos National Laboratory, Los Alamos, New Mexico 87545, USA\\
             $^2$Department of Physics, Sungkyunkwan University, Suwon 440-746,
             Korea\\
             $^3$Advanced Science Research Center, Japan Atomic Energy Agency, Tokai, Ibaraki 319-1195, Japan\\}

\date{\today}

\begin{abstract}
Heat capacity, magnetic susceptibility, NMR, and resistivity of
\Sr{} single crystals are presented, illustrating a purely
structural transition at 325 K with no magnetism. Bulk
superconductivity is found at 1.4 K. The magnitude of the transition
temperature $T_c$, fits to the heat capacity data, the small upper
critical field $H_{c2}$ = 390 Oe, and Ginzburg-Landau parameter
$\kappa$ = 2.1 suggests a conventional fully gapped superconductor.
With applied pressure  a second structural phase transition occurs which results in an 8\% reduction in the c/a ratio of lattice parameters. We find that superconductivity persists into
this high pressure phase, although the transition
temperature is monotonically suppressed with increasing pressure.
Comparison of these Ni-P data as well as layered Fe-As and Ni-As
superconductor indicates that reduced dimensionality can be a
mechanism for increasing the transition temperature.

\end{abstract}

\pacs{74.10.+v,74.25.Bt,74.70.Dd}

\maketitle
{\bf Introduction:}

There have long been attempts to identify structure-property
relations in superconducting materials that help in the
identification of the pairing mechanism and the enhancement of the
superconducting transition temperature. The discovery of
superconductivity at 26K in LaFeAsO\cite{KamiharaJACS2008} has
stimulated much recent work in compounds containing T-Pn layers
(T=transition metal, and Pn = pnictide). The highest transition
temperatures to date are found in the tetragonal ZrCuSiAs
structure-type (e.g. 55 K in SmFeAs(O,F)\cite{ZARen2008a}), and a
correlation has been identified between the As-Fe-As bond angle and
$T_c$ \cite{Lee2008JPSJ}. Clearly, structural tuning can have a very
dramatic effect on superconducting transition temperatures, and
further structure-property relations are needed to help guide the
discovery of higher transition temperatures.

Reasonably high transition temperatures also have been found in the
ThCr$_2$Si$_2$ structure (ie. 38 K in
(Ba,K)Fe$_2$As$_2$)\cite{Rotter2008b}. The \ThCrSi{} structure is stable for two different bonding configurations between the atoms in the Si position of neighboring Cr$_2$Si$_2$ planes. Either the atoms in the Si position are sufficiently far apart that they are in a non-bonding state, or they are close enough together that they are bonding (typical of single bond distances)\cite{HoffmannJPC1985}. Consequently, depending on the elements in the tetragonal \ThCrSi{} structure, one can induce an isostructural volume
collapse with either substitution or pressure
\cite{Huhnt1997,Keimes1997,Keimes1998,ShameemBanu1999}.
We will refer to the bonding configuration as the "collapsed
tetragonal" structure. Due to the increased bonding between layers,
it is natural to anticipate that the collapsed tetragonal phase will
be electronically more 3-dimensional (i.e. Fermi velocity ratio $v_x/v_z$ closer to 1) than the non-bonding
configuration. With Fe$_2$As$_2$ planes, the ThCr$_2$Si$_2$ structure
generally adopts a non-bonding configuration between As atoms from
neighboring planes. One exception is that \CaFeAs{} can be driven
into the collapsed tetragonal state with
pressure\cite{Kreyssig2008PRB}. Interestingly, while the homogeneous
collapsed tetragonal state does not show signs of superconductivity
down to 2 K\cite{Yu2008}, by using different pressure mediums a
structurally inhomogeneous scenario is induced \cite{Goldman2008}
which possesses an onset of superconductivity at 13 K
\cite{Park2008CaFe2As2, Torikachvili2008CaFe2As2}. This poses the
question of what is the role of the collapsed tetragonal phase (and
dimensionality in general) for superconductivity.

We attempt to shed light on this question with a study of
SrNi$_2$P$_2$. At ambient pressure, SrNi$_2$P$_2$ crystalizes in the
large tetragonal state above 325 K with an interlayer P-P distance of 3.120 \AA{} corresponding to a nonbonding condition\cite{Keimes1997}. Below 325 K, the system
undergoes an orthorhombic distortion (Immm space group) with a buckling in the Ni$_2$P$_2$ planes causing the interlayer P-P distance to oscillate between 3.282 \AA (nonbonding) and 2.452 \AA (bonding). With NMR measurements we show that there
is no magnetism associated with this transition. By applying
pressure, the collapsed tetragonal state can be induced with slight
pressure (4 kbar at room temperature)\cite{Keimes1997}. Here, we
first identify that SrNi$_2$P$_2$ is a bulk superconductor at
ambient pressure with $T_c$ = 1.4 K. An analysis of the data is
consistent with a fully gapped conventional superconductor. By
applying pressure, we find that the transition temperature is
suppressed by less than 50\% upon entering the collapsed tetragonal
state. Consequently, we conclude that the necessary conditions for superconductivity, namely the ability to create an attractive interaction which can overcome the screened coulomb repulsion between two quasiparticles, can be satisfied in the "collapsed tetragonal" state to the transition metal pnictides in the \ThCrSi{} structure. Furthermore, by expanding the cell so as to increase the c/a ratio superconductivity is enhanced.

{\bf Experimental:}

Large, plate-like single crystals of \Sr{} were grown in Sn flux
following the general recipe described in ref. \cite{Marchand1978}.
The starting materials in the ratio of Sr:Ni:P:Sn=1.3:2:2.3:16 were
placed in an alumina crucible and sealed under vacuum in a quartz
tube. The contents were then heated to 600 $^{\circ}$C for 4 hours
followed by 900 $^{\circ}$C for 200 hours. Subsequently the charge
was cooled to 650 $^{\circ}$C at a rate of 3.5 $^{\circ}$/hour at
which point the excess Sn was spun off with the aid of a centrifuge.
Powder X-ray diffraction confirm that the large plate like crystals
with typical dimensions 5 x 5 x 0.2 mm$^3$ were
\Sr\cite{Keimes1997}.

Specific heat measurements were performed using an adiabatic
relaxation technique in a Quantum Design PPMS. Resistance
measurements were also performed in a Quantum Design PPMS using a
Linear Research resistance bridge with an excitation current of 0.1
mA. Susceptibility measurements were performed in a Quantum Design
MPMS with an applied field of 5 T. Results were in good agreement
with data at 0.01 T, confirming that no ferromagnetic impurity
contribution exists in the sample. $^{31}$P NMR was performed using
a phase-coherent pulsed spectrometer. The external field was applied
along the c-axis.

For pressure measurements, a crystal was mounted inside a Teflon cup
within a hybrid BeCu/NiCrAl clamp-type pressure cell with silicon
oil as the pressure transmitting medium. The cup also contained a
small piece of Pb whose known pressure-dependent superconducting
transition \cite{Eiling1981} enabled a determination of the pressure
within the cell at low temperatures where the superconductivity is
measured.

{\bf Results:}

\begin{figure}
\includegraphics[width=3.3in]{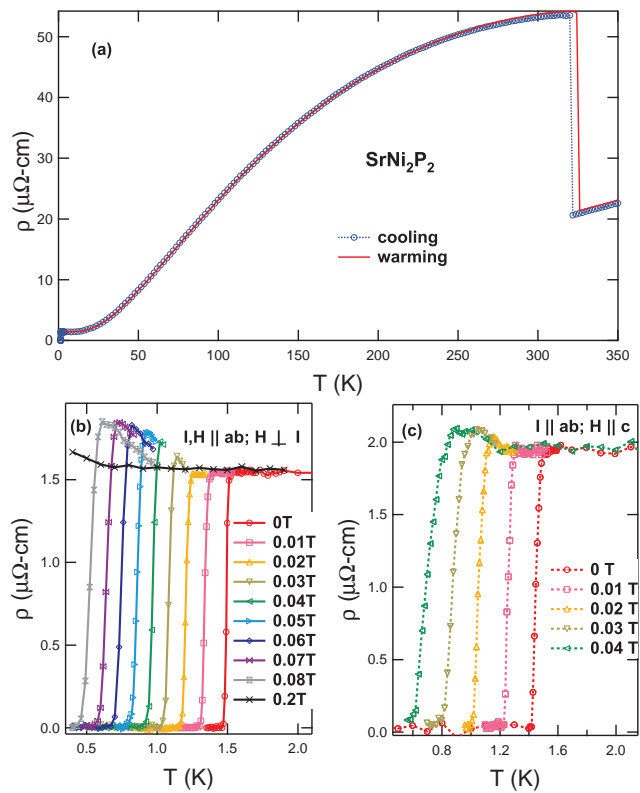}
\caption{(color online) In-plane electrical resistivity $\rho(T)$
(I$\parallel$ab) of \Sr{} (a) for cooling and warming through the
structural transition. (b) and (c) show the evolution of the
superconducting transition for $H
\parallel ab$ and $H \parallel c$, respectively.}
\label{Res}
\end{figure}

\begin{figure}
\includegraphics[width=3.3in]{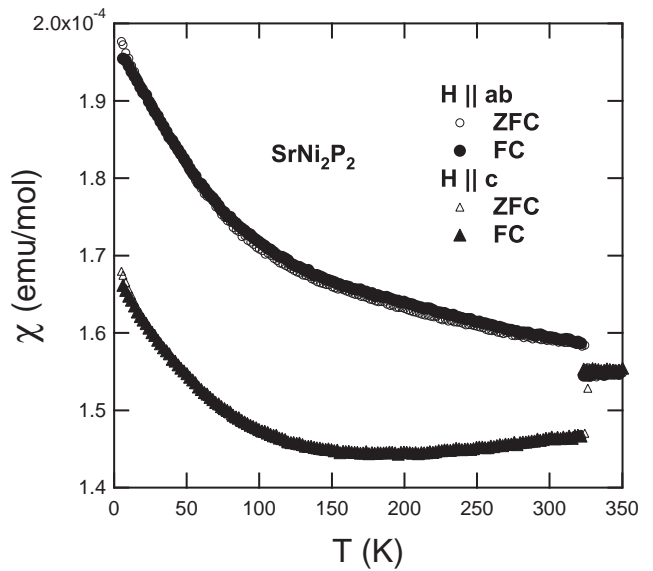}
\caption{Susceptibility of \Sr{} taken as a function of zero-field
cooled (ZFC) and field cooled (FC) with an applied field of 5 T.}
\label{Chi}
\end{figure}

The in-plane resistivity of \Sr{} is shown in figure \ref{Res}, and
demonstrates a sharp first order anomaly with thermal hysteresis
associated with the structural transition previously observed in
X-ray diffraction measurements\cite{Keimes1997}. Here we find the
transition occurs at 325 K upon warming and 320 K upon cooling. The
first order transition is further confirmed by magnetic
susceptibility measurements shown in figure \ref{Chi}. The
susceptibility also shows the first order transition at 325 K (upon
warming) and 322 K (upon cooling). Interestingly, the susceptibility
in the high temperature phase is isotropic, but below the transition
becomes anisotropic. While anisotropic g-factors can result in an
anisotropic Pauli susceptibility it is surprising that only the low temperature phase shows this anisotropy.

\begin{figure}
\includegraphics[width=3.3in]{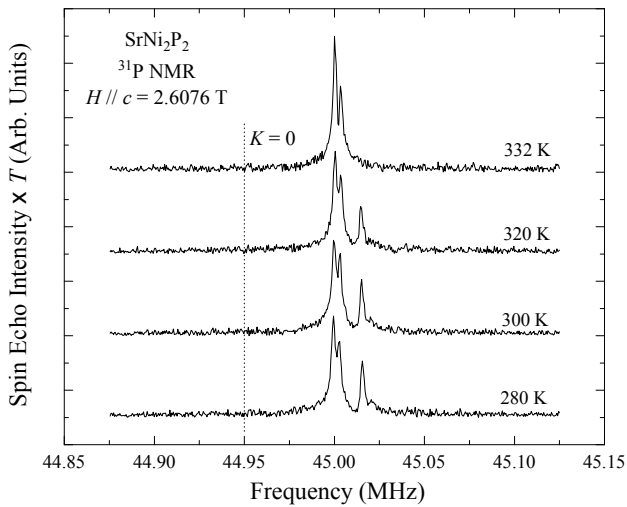}
\caption{$^{31}$P NMR of \Sr{} on cooling through the structural
transition illustrating the absence of magnetism. The dotted line
represents the resonance line in the absence of a Knight shift {\it K}.}
\label{NMR}
\end{figure}

To explore whether or not the susceptibility results could imply
magnetic ordering or a changing magnetic moment coincident with the
structural transition we performed $^{31}$P NMR above and below the
transition (see figure \ref{NMR}). The data above the transition
shows a very sharp double peak indicative of very high crystallinity.
Since only a single peak is anticipated in the high temperature
\ThCrSi{} structure, the double peak at 332 K may be a consequence
of either a slightly twinned sample, or a small structural
distortion possibly caused by thermally cycling through the
structural transition. Upon lowering temperature through the
transition, one sees that the central peak loses some intensity, and
a second double peak structure grows in at slightly higher
frequencies. This is as expected on the basis of the structural
transition alone for which there are now expected to be two
inequivalent P positions populated in a ratio of
2:1\cite{Keimes1997}. The fact that the central line does not shift
or broaden at all implies that there is no internal field on the P
sites which remain in the same crystallographic environment.
Consequently, we can conclude that there is no magnetic order or a
change in magnetic moment coincident with the structural transition
at 325 K.

The resistivity below the structural transition is characteristic of
a good metal with a RRR (= $\rho$(300K)$/\rho$(4K)) of 40 and a
residual resistivity of 1.8$\pm$0.2 $\mu\Omega$cm. At 1.45 K the
sample shows a sharp transition in resistivity which is consistent
with bulk superconductivity as established by heat capacity
measurements shown below. With an applied magnetic field the
superconducting transition is rapidly suppressed indicating a very
small upper critical field. In addition, with an applied magnetic
field there is a slight upturn in the resistivity before the
superconducting transition. A similar feature has been observed in
other iron-based pnictide superconductors and is discussed in ref.
\cite{Analytis2008LaFePO}.

\begin{figure}
\includegraphics[width=3.3in]{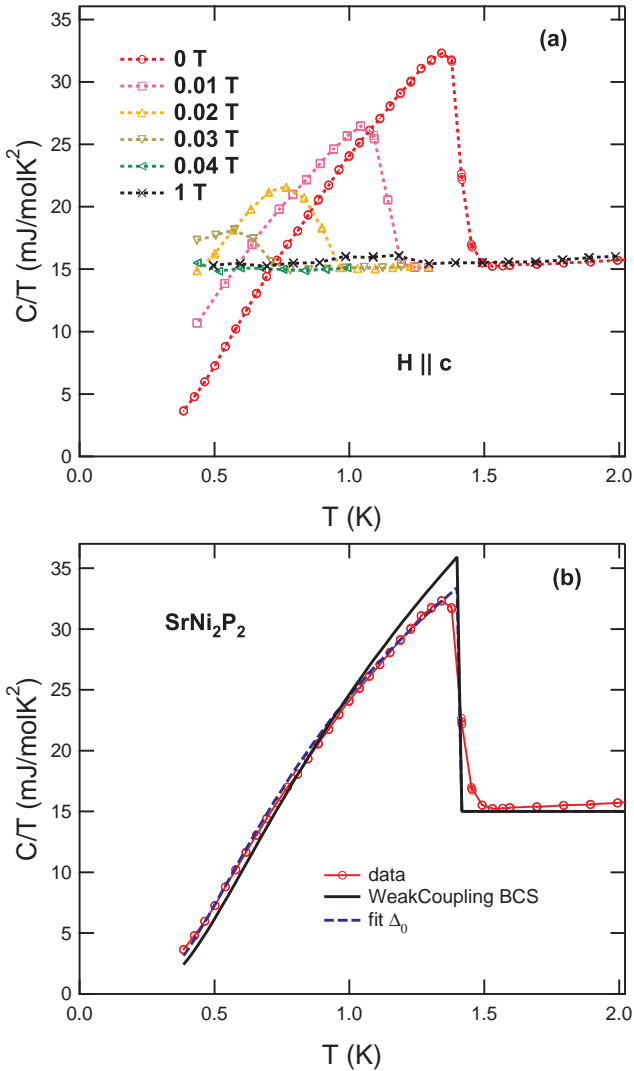}
\caption{(color online) (a) Specific heat versus temperature for
several magnetic fields applied in the plane of \Sr. (b) The zero
field heat capacity data (circles) along with the theoretical
expectation of the specific heat based on a purely weak coupling BCS
picture (solid black curve). Fit $\Delta_0$ (dashed blue line) is a
fit to the BCS expression holding $T_c$ = 1.415 K, $\gamma$ = 15
mJ/molK$^2$ constant as described in the text.} \label{Cp}
\end{figure}

Specific heat data at low temperatures are shown in figure \ref{Cp}.
A fit of the data (not shown) from 1.5 K to 10 K to $C/T = \gamma +
\beta T^2 + \delta T^4$ gives $\gamma$ = 15 mJ/mol K$^2$ and $\beta$
= 0.23 mJ/mol K$^4$, which implies a Debye temperature $\Theta_D$ of
348 K. Using a zero-temperature susceptibility of $\chi_0 =
(2\chi_{ab}$(2K) + $\chi_c$(2K))/3 = 1.86 emu/mol, we find a Wilson
Ratio $R_W$ of 0.9 using the expression $R_W$ =
$\pi^2k_B^2/3\mu_B^2(\chi_0/\gamma)$.

Bulk superconductivity is established by specific heat measurements.
By an equal area construction the specific heat gives a transition
temperature of 1.415 K, consistent with the drop in resistivity. The
fact that the low temperature specific heat gives a jump
$\Delta{}C/\gamma{}T_c$ = 1.27 confirms the bulk nature of
superconductivity. From BCS theory we expect the specific heat to be
given by the formula

\begin{eqnarray*}
C_\mathrm{BCS}=t\frac{d}{dt}\int_0^{\infty}\,dy(-\frac{6\gamma\Delta_0}{k_\mathrm{B}\pi^2})[f\ln{f}
+ (1-f)\ln{(1-f)}].
\end{eqnarray*}
where $t$\,=\,$T$/$T_\mathrm{c}$,
$f$\,=\,1/[$\exp$($E$/$k_\mathrm{B}$$T$)+1],
$E$\,=\,($\epsilon^2$+$\Delta^2$)$^{1/2}$,
$y$\,=\,$\epsilon$/$\Delta_0$, and $\Delta(T)/\Delta_0$ is taken
from the tables of M\"{u}hlschlegel \cite{Muhlschlegel}. The
specific heat curve, obtained by fixing $\gamma$ = 15 mJ/molK$^2$
from the high temperature fit, $T_c$ = 1.415 K and $\Delta_0$ to the
weak coupling BCS value = 1.76 $k_BT_c$, is shown by the solid black
curve in fig. \ref{Cp}b. Allowing the zero temperature gap value
$\Delta_0$ to vary we obtain the dashed blue curve and find
$\Delta_0$ = 0.201 meV = 1.65 $k_BT_c$. With the fitted curve to the
specific heat, we can extract the condensation energy and equate it
to the thermodynamic critical field giving $H_c$ = 185 Oe. The
reason that the best fit to the specific heat gives a value of
$\Delta_0$ {\it below} the weak coupling value is uncertain, but may
be caused by pair breaking impurities \cite{ParksSCtext}. We note
that in several related superconductors the specific heat jump is
also slightly smaller than the weak coupling BCS expectation of
$\Delta{}C/\gamma{}T_c$ = 1.43
\cite{RonningJPCM2008BaNi2As2,Bauer2008SrNi2As2,
Klimczuk2008La3Ni4P4O2,KasaharaJPCM2008,Kurita2008BaNi2As2Kappa}.

\begin{figure}
\includegraphics[width=3.3in]{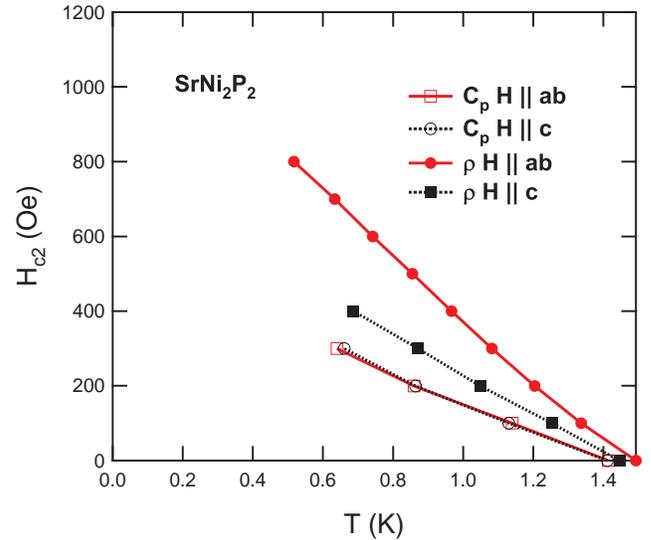}
\caption{(color online) Superconducting phase diagram of \Sr{} as
determined by the equal area construction of heat capacity data, and
from the midpoint of the resistive transition.} \label{HcvsTc}
\end{figure}

As mentioned above, the upper critical fields of \Sr{} are small. In
figure \ref{HcvsTc} we plot the upper critical field deduced from
the midpoint of the resistive transition as well as the midpoint of
the specific heat curves. There is a slight anisotropy observed for
the resistivity data. The upper critical field as determined by heat
capacity is slightly lower, but surprisingly, shows no anisotropy.
The slope of the upper critical field from heat capacity is
$dH_{c2}/dT_c$ = -0.039 T/K. This gives an upper critical field of
390 Oe using the WHH formula $H_{c2}^*$(0) =
-0.7$\,T_c\,dH_{c2}/dT_c$ \cite{WHH1966}, and a Ginzburg-Landau
coherence length $\xi_{GL}$ = 920 \AA{} from the expression
$H_{c2}(0) = \Phi_0/2 \pi \xi_{GL}^2$ where $\Phi_0$ = 2.07
10$^{-7}$ Oe cm$^2$ is the flux quantum. From the relations
$H_c/H_{c1} = H_{c2}/H_c = \surd2\kappa =
\surd2\lambda_{eff}/\xi_{GL}$ we extract $\kappa$ = 2.1, $H_{c1}$ =
88 Oe, and $\lambda_{eff} = 1935$ \AA.

\begin{figure}
\includegraphics[width=3.3in]{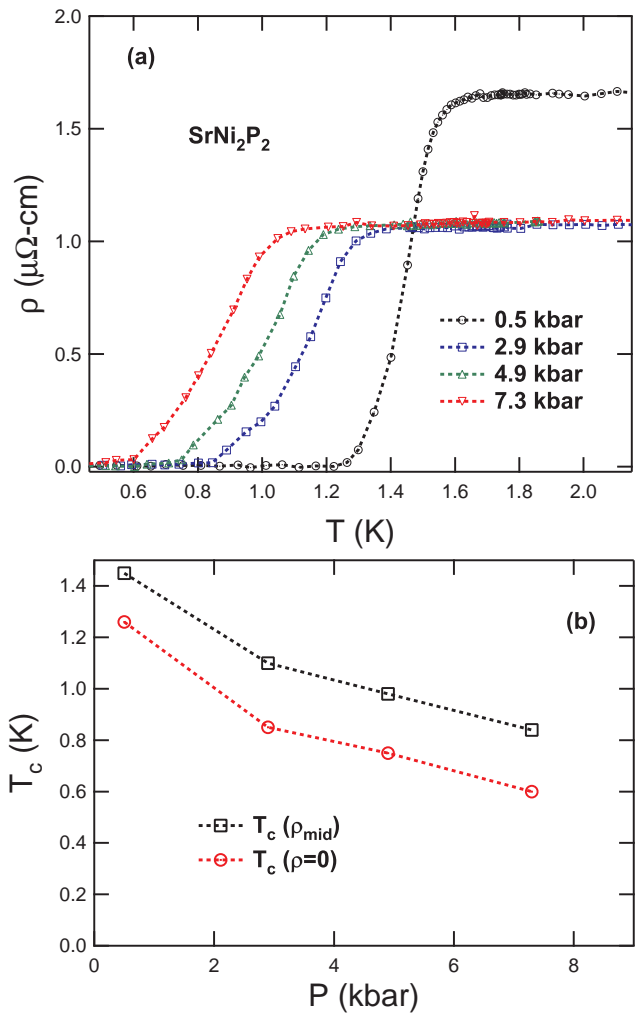}
\caption{(color online) (a) Resistivity versus temperature of \Sr{}
for several applied pressures. (b) $T_c$ as a function of pressure
as determined by the resistivity in (a) from both the midpoint of
the transition (black squares) and the zero resistance state (red
circles).} \label{Pressure}
\end{figure}

To investigate the influence of the crystal structure on
superconductivity, we applied pressure to \Sr. At room temperature
it is known that the orthorhombic structure transforms to a
collapsed tetragonal structure at 4 kbar\cite{Huhnt1997,Keimes1997}.
Based on the pressure-temperature phase diagram for the structural collapse in CaFe$_2$As$_2$\cite{Yu2008} we anticipate that at lower temperatures the structural transition to the "collapsed tetragonal" phase in \Sr{} will occur at pressures less than 4 kbar. The structural transformation results in
a volume change of -3.9\% and a -8\% change in the c/a
ratio\cite{Huhnt1997}. Low temperature resistivity is shown in
figure \ref{Pressure}a. There is an immediate discontinuous change
in the behavior from 0.5 kbar to 2.9 kbar. The change in normal
state value of the resistivity likely reflects the change in
structure which we anticipate with increasing pressure. The
superconducting transition temperature decreases monotonically with
pressure, although as shown in figure \ref{Pressure}b there is a
change in slope of $T_c$ versus pressure upon entering the high
pressure phase. (At 25 kbar (not shown) there was no
superconductivity observed down to 1.8 K.)  At pressures above 4 kbar we suggest the system is structurally homogeneous, and thus our results show that the collapsed tetragonal phase of transition-metal pnictides in the \ThCrSi{} structure can support superconductivity.

{\bf Discussion:}

Several results from the ambient pressure data of \Sr{} are
suggestive of a fully gapped BCS superconductor, including the fit
to the heat capacity data, the small value of $H_{c2}$ and $\kappa$.
Band structure calculations for \Sr{} give a density of states
$N(E_F)$ = 43 States/Ry cell \cite{ShameemBanu1999,
bandstructurefootnote}. This gives an electron-boson renormalization
$\lambda$ = 1.02, from the expression $\gamma = \pi^2k_B^2
N(E_F)(1+\lambda)/3$, using the experimentally measured $\gamma$ =
15 mJ/molK$^2$. This is certainly sufficient to explain
superconductivity in an electron-phonon pairing scenario. For
example, using the simplified McMillan formula \cite{McMillan1968}

\begin{eqnarray*}
T_\mathrm{c}=\frac{\Theta_D}{1.45}exp[-\frac{1.04(1+\lambda)}{\lambda-\mu^{\ast}(1+0.62\lambda)}]
\end{eqnarray*}

and a very conservative estimate of $\mu^{\ast}$ = 0.3 gives $T_c$ =
4.5 K. Furthermore, calculations show that \BaNiP{} has a larger
density of states ($\gamma^{th}$ = 9.32
mJ/molK$^2$)\cite{ShameemBanu1999} which is consistent with its
larger superconducting transition temperature ($T_c$ = 2.7
K)\cite{Mine2008BaNi2P2}. In addition, low temperature thermal
conductivity data provide compelling evidence for fully gapped
superconductivity in the related material
\BaNiAs\cite{Kurita2008BaNi2As2Kappa}. While not definitive, the
accumulation of evidence suggests that \Sr{} is a conventional BCS
superconductor.

An open question is whether the pairing mechanisms of the Fe based
systems and the Ni based compounds are related. On the basis of band
structure calculations it has been argued that both T = Fe and T =
Ni lie close to a magnetic instability \cite{Xu2008EPL}; however it
remains the case that none of the Ni based systems possess
superconducting transition temperatures in excess of 5
K\cite{Watanabe2007LaNiPO,Tegel2008LaNiPO,Mine2008BaNi2P2,FujiiJPCM2008,
Watanabe2008LaNiAsO,Fang2008NiAs,Li2008NiAs,RonningJPCM2008BaNi2As2,
Bauer2008SrNi2As2,Klimczuk2008La3Ni4P4O2,Ge2008GdNiBiO,Kozhevnikov2008LaNiBiO},
nor do they have evidence for magnetism. Density functional
calculations suggest that the two systems have unrelated pairing
mechanisms, based on the fact that they are unable to reproduce the
high transition temperatures of the Fe-based
compounds\cite{Sudedi2008LaNiPO}\cite{Sudedi2008BaNi2As2}. However,
it remains a possibility that the pairing mechanisms of the Ni-based
systems are merely not as optimized for superconductivity as their
iron-based cousins.

\begin{figure}
\includegraphics[width=3.3in]{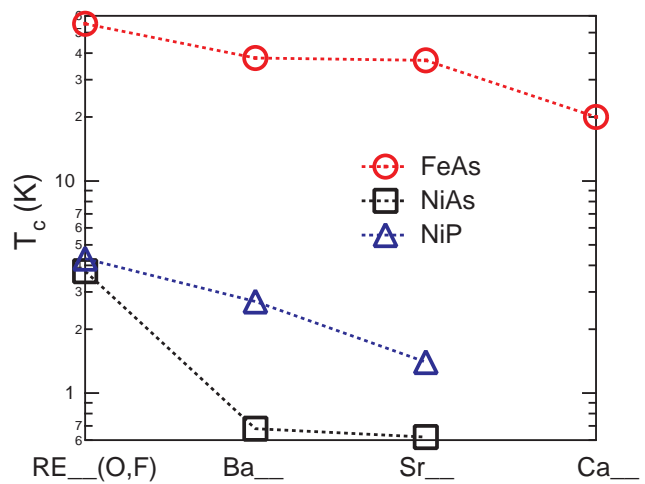}
\caption{(color online) (a) Superconducting transition temperature
across a few families of compounds for FeAs, NiAs, and NiP planes.
For the FeAs series, $T_c$ for SmFeAs(O,F)and for hole doping on the
Ba, Sr, Ca site was chosen (refs
\cite{ZARen2008a},\cite{Rotter2008b},\cite{Sasmal2008},\cite{Wu2008Ca}
respectively). $T_c$ for the NiP and NiAs were taken from refs.
\cite{Tegel2008LaNiPO},\cite{Mine2008BaNi2P2}, this work and
\cite{Fang2008NiAs}, \cite{RonningJPCM2008BaNi2As2}, and
\cite{Bauer2008SrNi2As2}, respectively.} \label{TcvsFamily}
\end{figure}

It is rather remarkable that superconductivity has been found in
many of the nickel analogs for which the iron-based pnictide
compounds also superconducts. Figure \ref{TcvsFamily} shows that
even the trend of the maximum superconducting transition temperature
decreases monotonically from ReX(O,F) to BaX$_2$ to SrX$_2$ to
CaX$_2$ irrespective of whether X = FeAs, NiAs, or NiP. A possible
interpretation of this trend is that $T_c$ is suppressed with
increasing dimensionality, or in other words, as the electronic structure becomes more isotropic. Comparison between the ZrCuSiAs and
\ThCrSi{} structure types has already established that the electronic structure of the systems in the \ThCrSi{} structure are more 3-dimensional than those in the ZrCuSiAs structure. \cite{Tanatar2009Ba122}. Within the ThCr$_2$Si$_2$
structure this can be further understood from the decreasing ionic
radius in going from Ba to Sr to Ca\cite{ShameemBanu1999,
HoffmannJPC1985}. Consequently, the pnictide atoms are closer
together allowing for stronger hybridization along the c-axis.
Eventually the decreasing pnictide-pnictide distance will lead to the collapsed tetragonal state where the interlayer pnictides adopt a bonding configuration, which
should be the most 3-dimensional due to the increased P-P
hybridization along the z-direction, and hence the least favorable
for superconductivity. Our results, which show that
superconductivity in SrNi$_2$P$_2$ is suppressed upon entering the
collapsed tetragonal phase, further support the notion that
increased 3-dimensionality found in the collapsed tetragonal state
at higher pressures is detrimental for superconductivity. Further
work is needed to determine how precisely dimensionality in terms of structural, magnetic, and electronic anisotropy  influences
the correlation between families observed in figure
\ref{TcvsFamily}. Irrespective of the origin of the trend, this work
suggests that CaNi$_2$P$_2$ and CaNi$_2$As$_2$ will be
superconducting at temperatures below 1.4 and 0.62 K, respectively.

{\bf Conclusion:}

In conclusion, by $^{31}$P NMR we have demonstrated the lack of
magnetism in \Sr{}. We have also shown that \Sr{} is a bulk
superconductor at ambient pressure at 1.4 K. Upon entering the
collapsed tetragonal phase with applied pressure, the
superconducting transition temperature is monotonically suppressed.
Further work is necessary to determine whether \Sr{} under pressure
is the first transition metal-pnictide in the \ThCrSi{} structure to
possess bulk superconductivity, or whether a similar unknown
mechanism as occurs in \CaFeAs{} is at play. The suppression of
$T_c$ with pressure and across families of compounds demonstrates
the notion that increased 2-dimensionality is a mechanism through
which higher $T_c$s can be achieved.

\begin{acknowledgments}
We thank M. Graf for helpful comments. Work at Los Alamos National
Laboratory was performed under the auspices of the U.S. Department
of Energy.
\end{acknowledgments}


\end{document}